\documentclass{article}

\PassOptionsToPackage{numbers, compress, square}{natbib}

\usepackage[final]{mlsb_2023}

\usepackage[utf8]{inputenc}
\usepackage[T1]{fontenc}
\usepackage{hyperref}
\usepackage{url}
\usepackage{booktabs}
\usepackage{amsfonts}
\usepackage{nicefrac}
\usepackage{microtype}
\usepackage{xcolor}
\usepackage{subfigure}
\usepackage{hyperref}
\usepackage{adjustbox}
\usepackage{amsmath}
\usepackage{amssymb}
\usepackage{mathtools}
\usepackage{amsthm}
\usepackage{svg}

\usepackage[capitalize,noabbrev]{cleveref}
\usepackage{wrapfig}

\usepackage{enumitem}

\definecolor{darkgreen}{rgb}{0.0, 0.5, 0.0}

\definecolor{UniBlue}{RGB}{0, 74, 153}
\definecolor{UniRed}{RGB}{193, 0, 42}
\definecolor{UniGrey}{RGB}{154, 155, 156}

\title{Rethinking Performance Measures of RNA Secondary Structure Problems}

\author{%
  Frederic Runge\\
  Department of Computer Science\\
  University of Freiburg\\
  \texttt{runget@cs.uni-freiburg.de} \\
  \And
  Jörg K.H. Franke \\
  Department of Computer Science\\
  University of Freiburg\\
  \texttt{frankej@cs.uni-freiburg.de} \\
  \And
  Daniel Fertmann \\
  Department of Computer Science\\
  University of Freiburg\\
  \texttt{fertmand@cs.uni-freiburg.de} \\
  \And
  Frank Hutter \\
  Department of Computer Science\\
  University of Freiburg\\
  \texttt{fh@cs.uni-freiburg.de} \\
}

\begin{document}

\maketitle

\begin{abstract}
Accurate RNA secondary structure prediction is vital for understanding cellular regulation and disease mechanisms. Deep learning (DL) methods have surpassed traditional algorithms by predicting complex features like pseudoknots and multi-interacting base pairs. However, traditional distance measures can hardly deal with such tertiary interactions and the currently used evaluation measures (F1 score, MCC) have limitations. We propose the Weisfeiler-Lehman graph kernel (WL) as an alternative metric. Embracing graph-based metrics like WL enables fair and accurate evaluation of RNA structure prediction algorithms. Further, WL provides informative guidance, as demonstrated in an RNA design experiment.
\end{abstract}

\section{Introduction}
\label{intro}

Ribonucleic acid (RNA) is one of the major regulators in cells and has been connected to multiple diseases like cancer~\citep{prensner_2011} and Parkinson's~\citep{cao_2018}.
Since the function of RNAs is dominated by their structure~\citep{gandhi_2018}, accurate prediction of these structures appears as a fundamental problem in computational biology~\citep{bonnet2020designing}.
RNA folds hierarchically and the formation of the final 3-dimensional shape strongly depends on the formation of a \emph{secondary structure}, which describes nucleotide pairings (base pairs) of the RNA sequence via hydrogen bonds~\citep{tinoco_1999}. 
The secondary structure already defines the sites for interactions with other cellular compounds~\citep{gandhi_2018}, and improvements in secondary structure prediction, therefore, could have a substantial impact on RNA-related research.

The potential improvements in the field of RNA structure prediction by sophisticated learning algorithms recently attracted the interest of the deep learning (DL) community and led to an explosion of DL-based approaches in the field~\citep{singh2019spotrna, zhang2019new, rezaur2019learning, singh2021spotrna2, saman2022rna, wayment2022rna, jungrtfold, franke2022probabilistic, chen2022rna_fm, chen2023redfold,franke2023scalable}, and new state-of-the-art results.
The typical output of these algorithms is a squared $L\times L$ binary adjacency matrix, where $L$ is the length of the input nucleotide sequence, indicating positions where base pairs form~\citep{singh2019spotrna, franke2023scalable}.
Therefore, they can predict all possible base pairs, including pseudoknots~\citep{staple2005pseudoknots} and multi-interacting bases (multiplets) which is an advantage over more traditional, dynamic programming-based algorithms.

\begin{figure*}[t]
\begin{center}
        \resizebox{0.9\linewidth}{!}{%
        \centerline{%
        \includegraphics{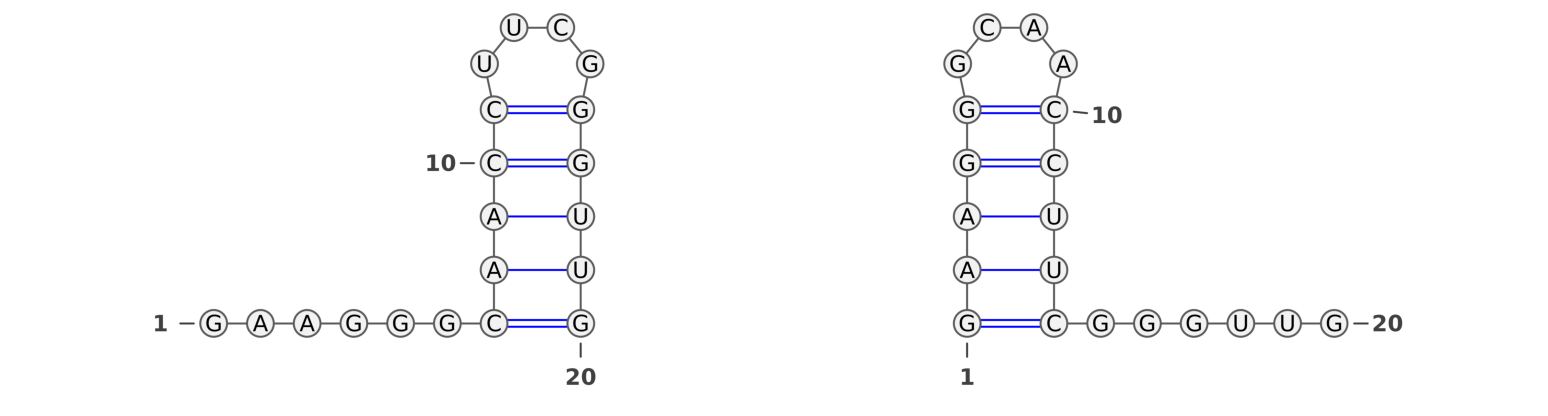}%
        }}
\caption[skip=0.1in]{
    \textbf{Bi-stable RNA 20mer.} The F1 score and MCC when comparing both folds is $0.0$ and $-0.026$, respectively. The Weisfeiler-Lehman graph kernel provides a score of $0.25$.
	}	\label{fig:mcc_f1_fail}
\end{center}
\vskip -0.2in
\end{figure*}

RNA secondary structure predictions are typically evaluated for the accuracy of predicted base pairs compared to a known structure.
Traditional algorithms typically use tree representations, either defining structure distance via edit operations~\citep{zhang1989simple, klein1998computing, demaine2009optimal} or tree alignment approaches~\citep{hochsmann2003local}.
However, while there exist methods that are capable of comparing pseudoknotted structures~\citep{quadrini2019algebraic}, to the best of our knowledge, there exists currently no algorithm that considers base multiplets, an important type of nucleotide interactions e.g. for the formation of G-quadruplex structures~\citep{gellert1962helix}. 
Therefore, current state-of-the-art DL approaches evaluate their predictions using performance measures derived from the confusion matrix.
Two performance measures are well established, the F1 score and the Matthews Correlation Coefficient (MCC)~\citep{matthews1975comparison}.
The main difference is that the F1 score is independent of the true negatives and that it is not symmetric with respect to class swapping which makes the MCC generally preferable with respect to binary classification~\citep{chicco2020advantages}.
However, for scoring secondary structure prediction, both have their flaws.
As an example, Figure~\ref{fig:mcc_f1_fail} shows a bi-stable RNA 20mer taken from \citet{wenter2006caged} in its two conformations. The respective F1 score between both conformations is zero, while the MCC score of $-0.026$ is even below the score expected for a random structure. 
Obviously, both structures share a common feature (the hairpin structure) with the exact same base pair pattern which should be reflected by the score.
\citet{mathews2019benchmark} recently proposed a derivative of the F1 score, named F1-shift for the remainder of this paper, to account for RNA structural dynamics like bulge migration~\citep{woodson1987flucuatingpairs}.
While this measure accounts for certain shifts in the base pairing scheme of the secondary structure, it still cannot capture the similarity between secondary structures as shown in Figure~\ref{fig:mcc_f1_fail}, similarly resulting in a score of zero.

RNAs can be represented as graphs and, therefore, graph metrics could be considered for scoring the distance between two RNA secondary structures.
To our knowledge, graph distance metrics have not yet been used for scoring RNA secondary structure prediction.
Hence, we propose to use the Weisfeiler-Lehman graph kernel~\citep{shervashidze2011weisfeiler} for more accurate evaluations in RNA secondary structure prediction tasks, instead of the commonly used F1 score, MCC, or their derivatives for binary classification.
In particular, our main contributions are as follows:

\begin{itemize}[label=$\bullet$, nosep, leftmargin=*, noitemsep, before={\vspace*{-0.25\baselineskip}}]
  \item We show that the commonly used F1 score and MCC as metrics for evaluating the quality of the secondary structure are misleading and error-prone.
  \item We propose the Weisfeiler-Lehman graph kernel (WL) as an alternative metric to closer align the biological motivation with the metric scale.
  \item We provide real-world examples to showcase the benefit of WL evaluation in contrast to F1 score, shifted F1 score, and MCC.
  \item In two application settings, we show the practical usefulness of WL for analysis and to guide an RNA design algorithm with improved results compared to traditional distance measures.
\end{itemize}

\section{RNA Secondary Structure Measures}

We review the commonly used measures for RNA secondary structure prediction, F1 score, MCC, and F1-shift before we briefly introduce the Weisfeiler-Lehman (WL) kernel, a powerful graph similarity measure widely used for comparing graphs in various domains. The WL kernel leverages the concept of graph isomorphism and employs a label propagation approach to capture the structural information of graphs.

\subsection{Confusion Matrix Based Measures}
The commonly used performance measures for RNA secondary structure prediction are based on a confusion matrix, which describes the number of true positives (TP), true negatives (TN), false positives (FP), and false negatives (FN) of a given prediction.

\paragraph{F1 Score}
The F1 score describes the harmonic mean of precision and recall and can be described as $F1 = 2 \cdot TP / (2 \cdot TP + FP + FN)$.

\paragraph{Matthews Correlation Coefficient}
While the F1 score emphasizes on positives, the MCC is a more balanced measure. The MCC can be calculated as follows.

\begin{equation}
   MCC = \frac{(TP \cdot TN) - (FP \cdot FN)}{\sqrt{(TP + FP) \cdot (TP + FN) \cdot (TN + FP) \cdot (TN + FN)}}
\end{equation}

\paragraph{F1-shift} 
The F1-shift is a measure to account for structural dynamics in RNAs~\citep{mathews2019benchmark}. 
The F1-shift is computed as the F1 score, but for a given pair $(i,j)$ all pairs $(i,j+1)$, $(i+1,j)$, $(i,j-1)$, and $(i-1,j)$ are also considered correct.

\subsection{Graph Isomorphism}
Graph isomorphism refers to the notion of two graphs being structurally identical. Given two graphs $G_1=(V_1, E_1)$ and $G_2=(V_2, E_2)$, an isomorphism between them is a bijective mapping $f: V_1 \rightarrow V_2$ that preserves the adjacency relationship between vertices. Formally, for any two vertices $u, v \in V_1$, $(u, v) \in E_1$ if and only if $(f(u), f(v)) \in E_2$. Determining graph isomorphism is a computationally challenging problem with significant implications in various domains.

\subsection{The Weisfeiler-Lehman Kernel}
The Weisfeiler-Lehman kernel is a graph kernel that captures the structural information of graphs by iteratively refining node labels based on their local neighborhoods. It operates in two main steps: label propagation and hash function computation. The kernel assigns each node in the graph a label representing the node's local structural information and then computes a hash function that aggregates these labels to generate a feature vector.

\subsubsection{Label Propagation}

The label propagation step involves iterating over the nodes of the graph and updating their labels based on the labels of their neighboring nodes. Initially, each node is assigned a unique label. In each iteration, the kernel collects the labels of a node's neighbors, sorts them lexicographically, and appends the node's own label to the list. This combined list of labels serves as the input for a hash function in the next step.
Let $L_i(u)$ denote the label of node $u$ at iteration $i$. The label propagation process can be defined as follows:

\begin{equation}
    L_{i+1}(u) = \text{hash}\left(L_i(u), \text{sort}\left(\left(L_i(v_1), \dots, L_i(v_{\vert N(u) \vert})\right)\right)\right), v_j \in N(u), j \in \{1, \dots, \vert N(u) \vert\},
\end{equation}

where $N(u)$ represents the set of neighboring nodes of $u$, $\text{sort}(\cdot)$ sorts the labels lexicographically, and $\text{hash}(\cdot)$ computes a hash value.

\subsubsection{Hash Function Computation}

The hash function computes a hash value based on the lexicographically sorted list of labels obtained from the label propagation step. This hash value captures the local structural information of a node's neighborhood and is used to refine the node's label in the subsequent iterations. The hash function is typically implemented using a simple and efficient algorithm, such as the Rabin-Karp hash function~\citep{karp1987efficient}.
The WL kernel computes a similarity score between two graphs $G_1$ and $G_2$ by comparing their corresponding label distributions obtained from the label propagation process. The similarity score can be computed as the dot product of the feature vectors generated by the hash functions:

\begin{equation}
    \text{WL-Similarity}(G_1, G_2) = \Phi(G_1) \cdot \Phi(G_2),
\end{equation}

where $\Phi(G)$ represents the feature vector of graph $G$ obtained by aggregating the labels through the hash functions.

The WL kernel is a powerful metric for comparing graphs, as it captures the structural information while being computationally efficient. It has been widely adopted in graph classification, pattern recognition, and graph mining tasks, showcasing its effectiveness across various domains.
In the following, we use five iterations of WL for all numbers reported. 

\begin{table*}[t]
\centering
\caption{Evaluation of RNA secondary structure prediction algorithms. Note that we are not interested in finding the state-of-the-art algorithm but in a comparison of the different performance measures.}
\label{tbl:benchmark}
\resizebox{1.0\linewidth}{!}{
\begin{tabular}{@{}lrrrrrrrr@{}}
\toprule
Model & F1 & F1-shift & MCC & WL & F1 Rank & F1-shift Rank & MCC Rank & WL Rank \\ \midrule
RNAformer      & 0.712 & 0.721 & 0.731 & 0.771 & 1 & 1 & 1 & 1 \\
SPOT-RNA       & 0.672 & 0.691 & 0.687 & 0.707 & 2 & \color{UniRed}3\color{black} & 2 & \color{UniRed}3\color{black} \\
SPOT-RNA2      & 0.668 & 0.701 & 0.674 & 0.705 & 3 & \color{UniRed}2\color{black} & \color{UniRed}4\color{black} & \color{UniRed}4\color{black} \\
RNA-FM         & 0.665 & 0.691 & 0.683 & 0.709 & 4 & \color{UniRed}3\color{black} & \color{UniRed}3\color{black} & \color{UniRed}2\color{black} \\
RNAFold        & 0.636 & 0.659 & 0.641 & 0.695 & 5 & \color{UniRed}4\color{black} & 5 & 5 \\
LinearFold-V   & 0.633 & 0.659 & 0.638 & 0.694 & 6 & \color{UniRed}4\color{black} & 6 & 6 \\
ContraFold     & 0.625 & 0.659 & 0.636 & 0.692 & 7 & \color{UniRed}4\color{black} & 7 & 7 \\
pKiss          & 0.613 & 0.64  & 0.62  & 0.676 & 8 & \color{UniRed}5\color{black} & \color{UniRed}10\color{black} & \color{UniRed}9\color{black} \\
ipKnot         & 0.611 & 0.617 & 0.624 & 0.675 & 9 & \color{UniRed}8\color{black}& 9 & \color{UniRed}10\color{black} \\
LinearFold-C   & 0.61  & 0.63  & 0.628 & 0.686 & 10 & \color{UniRed}7\color{black} & \color{UniRed}8\color{black} & \color{UniRed}8\color{black} \\
RNAstructure   & 0.606 & 0.633 & 0.611 & 0.667 & 11& \color{UniRed}6\color{black} &11 & 11\\
REDfold        & 0.487 & 0.502 & 0.501 & 0.619 & 12& \color{UniRed}9\color{black}& 12 & 12\\
\bottomrule
\end{tabular}%
}
\end{table*}

\begin{figure*}[t]
\vspace{-5pt}
\begin{center}
        \centerline{%
        \includegraphics[width=\textwidth,]{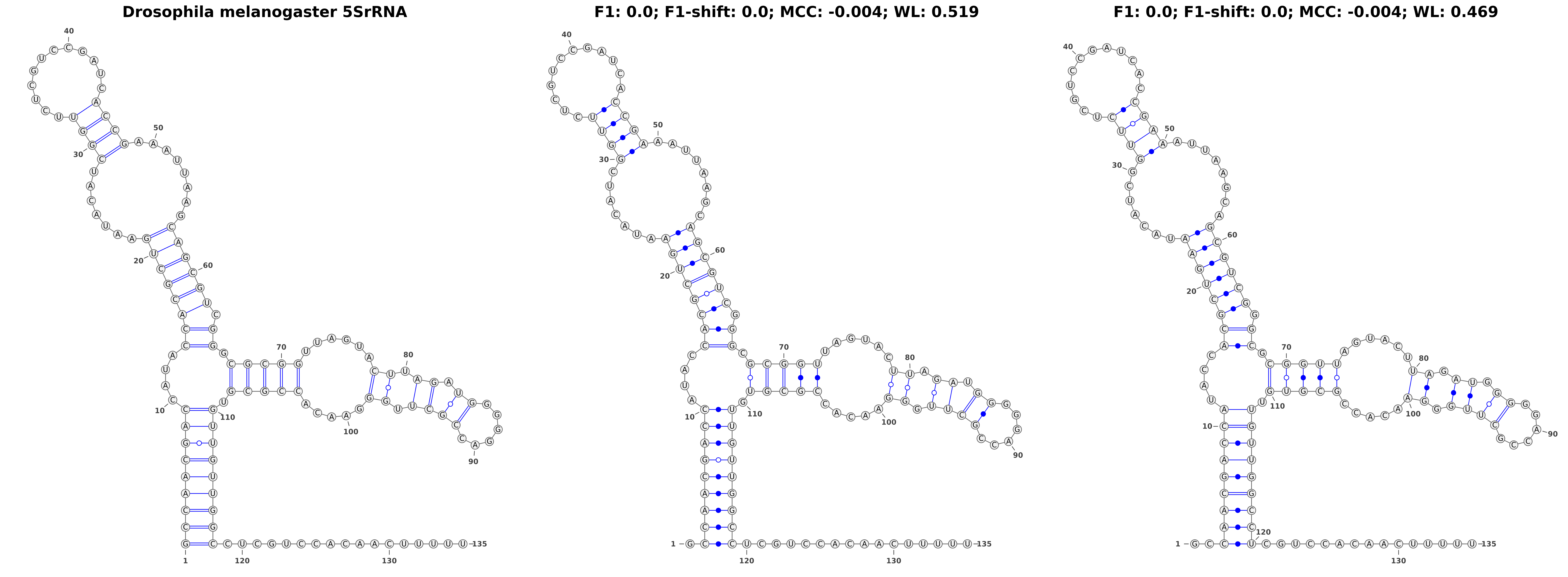}%
        }%}
\caption{
    \textbf{Example of structural shift.} (Left) We show a 5SrRNA of Drosophila melanogaster (Middle) The same structure shifted by one position. (Right) The same structure shifted by two positions.
	}	\label{fig:shift_example}
\end{center}
\vskip -0.2in
\end{figure*}

\section{The Weisfeiler-Lehman Kernel for RNA Secondary Structure Measure}
\label{discussion}
Before discussing potential misleading properties of the currently used performance measures for RNA secondary structure prediction, F1 score, F1-shift, and MCC, and displaying the benefits of WL, we start our discussion with an overview of the results, given by the different measures.
\paragraph{Evaluation of RNA Folding Algorithms} We evaluate several RNA secondary structure prediction algorithms\citep{hofacker_1994, do2006contrafold, reuter2010rnastructure, sato2011ipknot, janssen2015pkiss, huang2019linearfold, singh2019spotrna, singh2021spotrna2, chen2022rna_fm, chen2023redfold, franke2023scalable} on the most recently proposed benchmark dataset for RNA secondary structure prediction, TS-hard~\citep{singh2021spotrna2}. 
The TS-hard test set is derived from 3D structures of the Protein Data Bank (PDB)~\citep{wwpdb_2019} and contains a total of 28 samples; 7 samples without pseudoknots and nucleotides that pair with more than one other nucleotide (multiplets), 1 sample without pseudoknots but with multiplets, and 20 samples with both, pseudoknots and multiplets. 
We rank all algorithms with respect to the different performance measures.
Table~\ref{tbl:benchmark} shows the results of our evaluation.
We observe that the choice of the performance measure can change the order of algorithms drastically. 

\paragraph{Properties of Performance Measures} We continue with analyzing, what we believe to be, a misleading property of the currently used performance measures F1 score, F1-shift, and MCC while showing that the WL approach can provide more informative scores for the specific setting. 
As an example, we use the secondary structure of a 5SrRNA of \emph{Drosophila melanogaster} as provided by RNAcentral~\citep{rnacentral2021rnacentral} (RNAcentral ID: URS00003B4856\_7227).
We then introduce a shift in the structure by one and two positions.
As shown in Figure~\ref{fig:shift_example}, the structures all look similar, except for the positional shift. 
However, due to the binary nature of the scores, the F1 score as well as the F1-shift drop to zero for the shifted structures, while the MCC even shows a negative score.
Further, the scores remain unchanged between the one positional and two positional shifts.
In contrast, the WL captures the shift very well, resulting in a score of $0.519$ for the single positional shift and $0.469$ for the shift by two positions. Similarly, we simulate a bulge migration process, shown in Figure~\ref{fig:bulge_migration} in Appendix~\ref{app:bulge_migration}. The F1-shift measure was introduced to capture such events and shows a score of $1.0$ for all structures. However, there exist structures that do not show the migrating phenomenon. For example the HIV-1 TAR RNA shows a specific stabilized tri-nucleotide bulge that is essential for Tat protein binding and obligatory in virus replication~\citep{roy1990bulge, kulinski2003apical, chen2018exosomes}. Thus, we think that a score should still be able to quantify a difference between the structures. We observe that the score of WL gradually decreases with the bulge moving further away from the original point similar to F1 score and MCC, however, the decrease is smaller, and less strict than for the other measures. 
We believe that these are very valuable and desirable properties of WL, which we can also leverage during training:
While all of the measures, including WL, are not differentiable, there exist surrogate models capable of accurately modeling graph distances based on graph neural networks (GNNs)~\citep{damke2020novel}. Such a surrogate model could be used during training to inform a learning algorithm for secondary structure prediction while being fully differentiable. 

\begin{wrapfigure}{r}{0.5\linewidth}
\vspace{-24pt}
\begin{center}
        \resizebox{\linewidth}{!}{%
        \includegraphics[width=\textwidth]{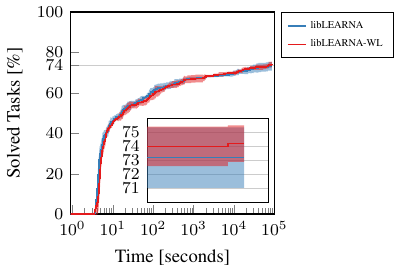}%
        }%}
\end{center}
\caption{RNA Design guided by Hamming distance (libLEARNA) or WL (libLEARNA-WL).}
\label{fig:eterna100}
\vspace{-10pt}
\end{wrapfigure}

\paragraph{Evolutionary Distance} A strong advantage of WL is the inclusion of sequence information into structure evaluation. We demonstrate this by simulating mutation events on the sequence level. The results are shown in Figure~\ref{fig:mutation} in Appendix~\ref{app:mutation}. While all other measures cannot capture the mutation information, the WL decreases with the amount of sequence changes. This could be a useful feature of WL e.g. when applying it in evolutionary studies to determine distances on the sequence and structure level.

\paragraph{Application to RNA Design} We see another application of WL in the field of RNA design. Here, we use WL to guide the design of the most recently proposed learning-based algorithm, libLEARNA~\citep{runge2023towards}, an improved version of the automated reinforcement learning approach, LEARNA~\citep{runge_2019}. Specifically, we use a model that was trained using the Hamming distance to measure the distance between the target structure and the folding of the predicted candidate sequence and exchange this distance measure with WL during evaluation. We use version two~\citep{koodli2021redesigning} of the commonly used Eterna100 benchmark~\citep{anderson-lee_2016}, with the proposed evaluation scheme of five independent runs for 24 hours. Figure~\ref{fig:eterna100} shows that without training on it, WL seems to improve the guidance of libLEARNA during evaluation, resulting in improved performance.

\paragraph{Limitations} 
A common phenomenon in RNA is base-stacking~\citep{ali2023structural}, which cannot be captured by any of the performance measures compared here. Further, while the WL allows to include sequence information which makes it aware of changes in the base pairing pattern, vanilla WL is unaware of the value of exchanging specific base pairs (e.g. Watson-Crick and non-canonical base pairs). For example, it might be desirable to penalize the introduction of certain base interactions (e.g. pseudoknots) when these are not formed in the ground-truth structure. However, both, the base-stacking and base pair penalties, could generally be introduced via weights of edges to inform the WL about such changes, but such an adapted version of WL demands in-depth analysis in the future. 

\section{Conclusion}
\label{conclusion}
We show that currently used performance measures for the evaluation of RNA secondary structure prediction have misleading properties due to their focus on the confusion matrix.
We propose to use graph distance metrics for the evaluation of the secondary structure prediction and suggest the Weisfeiler-Lehman graph kernel (WL) as a competent measure of graph similarity.
Subsequently, we compare the Weisfeiler-Lehman graph kernel to current measures in different settings, indicating its benefits and limitations.
Finally, we suggest that GNN-based surrogate models can be used to train DL algorithms more informed for RNA secondary structure prediction and show that WL can improve the performance when applied to RNA design.

\begin{ack}
The authors acknowledge support by the state of Baden-W\"{u}rttemberg through bwHPC and the German Research Foundation (DFG) through grant no INST 39/963-1 FUGG.
\end{ack}

\bibliographystyle{unsrtnat}
\bibliography{rna,rna_new,rna_data,rna_folding,dl,rna_design}

\begin{thebibliography}{50}
\providecommand{\natexlab}[1]{#1}
\providecommand{\url}[1]{\texttt{#1}}
\expandafter\ifx\csname urlstyle\endcsname\relax
  \providecommand{\doi}[1]{doi: #1}\else
  \providecommand{\doi}{doi: \begingroup \urlstyle{rm}\Url}\fi

\bibitem[Prensner et~al.(2011)Prensner, Iyer, Balbin, Dhanasekaran, Cao,
  Brenner, Laxman, Asangani, Grasso, Kominsky, Cao, Jing, Wang, Siddiqui, Wei,
  Robinson, Iyer, Palanisamy, Maher, and Chinnaiyan]{prensner_2011}
John~R. Prensner, Matthew~K. Iyer, O.~Alejandro Balbin, Saravana~M.
  Dhanasekaran, Qi~Cao, J.~Chad Brenner, Bharathi Laxman, Irfan~A. Asangani,
  Catherine~S. Grasso, Hal~D. Kominsky, Xuhong Cao, Xiaojun Jing, Xiaoju Wang,
  Javed Siddiqui, John~T. Wei, Daniel Robinson, Hari~K. Iyer, Nallasivam
  Palanisamy, Christopher~A. Maher, and Arul~M. Chinnaiyan.
\newblock Transcriptome sequencing across a prostate cancer cohort identifies
  pcat-1, an unannotated lincrna implicated in disease progression.
\newblock \emph{Nature Biotechnology}, 29\penalty0 (8):\penalty0 742--749,
  2011.

\bibitem[Cao et~al.(2018)Cao, Wang, Qu, Kang, and Yang]{cao_2018}
Bingqing Cao, Tao Wang, Qiumin Qu, Tao Kang, and Qian Yang.
\newblock Long noncoding rna snhg1 promotes neuroinflammation in parkinson’s
  disease via regulating mir-7/nlrp3 pathway.
\newblock \emph{Neuroscience}, 388:\penalty0 118 -- 127, 2018.
\newblock ISSN 0306-4522.

\bibitem[Gandhi et~al.(2018)Gandhi, Caudron-Herger, and
  Diederichs]{gandhi_2018}
Minakshi Gandhi, Maiwen Caudron-Herger, and Sven Diederichs.
\newblock Rna motifs and combinatorial prediction of interactions, stability
  and localization of noncoding rnas.
\newblock \emph{Nature Structural \& Molecular Biology}, 25:\penalty0
  1070--1076, 2018.

\bibitem[Bonnet et~al.(2020)Bonnet, Rz{a}{\.{z}}ewski, and
  Sikora]{bonnet2020designing}
{\'E}douard Bonnet, Pawe{\l} Rz{a}{\.{z}}ewski, and Florian Sikora.
\newblock Designing rna secondary structures is hard.
\newblock \emph{Journal of Computational Biology}, 27\penalty0 (3):\penalty0
  302--316, 2020.

\bibitem[Tinoco~Jr and Bustamante(1999)]{tinoco_1999}
Ignacio Tinoco~Jr and Carlos Bustamante.
\newblock How rna folds.
\newblock \emph{Journal of molecular biology}, 293\penalty0 (2):\penalty0
  271--281, 1999.

\bibitem[Singh et~al.(2019)Singh, Hanson, Paliwal, and Zhou]{singh2019spotrna}
Jaswinder Singh, Jack Hanson, Kuldip Paliwal, and Yaoqi Zhou.
\newblock Rna secondary structure prediction using an ensemble of
  two-dimensional deep neural networks and transfer learning.
\newblock \emph{Nature communications}, 10\penalty0 (1):\penalty0 1--13, 2019.

\bibitem[Zhang et~al.(2019)Zhang, Zhang, Li, Li, Wei, Zhang, and
  Liu]{zhang2019new}
Hao Zhang, Chunhe Zhang, Zhi Li, Cong Li, Xu~Wei, Borui Zhang, and Yuanning
  Liu.
\newblock A new method of rna secondary structure prediction based on
  convolutional neural network and dynamic programming.
\newblock \emph{Frontiers in genetics}, 10:\penalty0 467, 2019.

\bibitem[Rezaur Rahman~Chowdhury et~al.(2019)Rezaur Rahman~Chowdhury, Zhang,
  and Huang]{rezaur2019learning}
FA~Rezaur Rahman~Chowdhury, He~Zhang, and Liang Huang.
\newblock Learning to fold rnas in linear time.
\newblock \emph{bioRxiv}, page 852871, 2019.

\bibitem[Singh et~al.(2021)Singh, Paliwal, Zhang, Singh, Litfin, and
  Zhou]{singh2021spotrna2}
Jaswinder Singh, Kuldip Paliwal, Tongchuan Zhang, Jaspreet Singh, Thomas
  Litfin, and Yaoqi Zhou.
\newblock Improved rna secondary structure and tertiary base-pairing prediction
  using evolutionary profile, mutational coupling and two-dimensional transfer
  learning.
\newblock \emph{Bioinformatics}, 37, 2021.

\bibitem[Saman~Booy et~al.(2022)Saman~Booy, Ilin, and Orponen]{saman2022rna}
Mehdi Saman~Booy, Alexander Ilin, and Pekka Orponen.
\newblock Rna secondary structure prediction with convolutional neural
  networks.
\newblock \emph{BMC bioinformatics}, 23\penalty0 (1):\penalty0 58, 2022.

\bibitem[Wayment-Steele et~al.(2022)Wayment-Steele, Kladwang, Strom, Lee,
  Treuille, Becka, Participants, and Das]{wayment2022rna}
Hannah~K Wayment-Steele, Wipapat Kladwang, Alexandra~I Strom, Jeehyung Lee,
  Adrien Treuille, Alex Becka, Eterna Participants, and Rhiju Das.
\newblock Rna secondary structure packages evaluated and improved by
  high-throughput experiments.
\newblock \emph{Nature Methods}, 19\penalty0 (10):\penalty0 1234--1242, 2022.

\bibitem[Jung et~al.()Jung, Lee, Gao, and Frey]{jungrtfold}
Andrew~J Jung, Leo~J Lee, Alice~J Gao, and Brendan~J Frey.
\newblock Rtfold: Rna secondary structure prediction using deep learning with
  domain inductive bias.

\bibitem[Franke et~al.(2022)Franke, Runge, and Hutter]{franke2022probabilistic}
J{\"o}rg Franke, Frederic Runge, and Frank Hutter.
\newblock Probabilistic transformer: Modelling ambiguities and distributions
  for rna folding and molecule design.
\newblock \emph{Advances in Neural Information Processing Systems},
  35:\penalty0 26856--26873, 2022.

\bibitem[Chen et~al.(2022)Chen, Hu, Sun, Tan, Wang, Yu, Zong, Hong, Xiao, King,
  et~al.]{chen2022rna_fm}
Jiayang Chen, Zhihang Hu, Siqi Sun, Qingxiong Tan, Yixuan Wang, Qinze Yu,
  Licheng Zong, Liang Hong, Jin Xiao, Irwin King, et~al.
\newblock Interpretable rna foundation model from unannotated data for highly
  accurate rna structure and function predictions.
\newblock \emph{arXiv preprint arXiv:2204.00300}, 2022.

\bibitem[Chen and Chan(2023)]{chen2023redfold}
Chun-Chi Chen and Yi-Ming Chan.
\newblock Redfold: accurate rna secondary structure prediction using residual
  encoder-decoder network.
\newblock \emph{BMC bioinformatics}, 24\penalty0 (1):\penalty0 1--13, 2023.

\bibitem[Franke et~al.(2023)Franke, Runge, and Hutter]{franke2023scalable}
J{\"o}rg~KH Franke, Frederic Runge, and Frank Hutter.
\newblock Scalable deep learning for rna secondary structure prediction.
\newblock \emph{arXiv preprint arXiv:2307.10073}, 2023.

\bibitem[Staple and Butcher(2005)]{staple2005pseudoknots}
David~W Staple and Samuel~E Butcher.
\newblock Pseudoknots: Rna structures with diverse functions.
\newblock \emph{PLoS biology}, 3\penalty0 (6):\penalty0 e213, 2005.

\bibitem[Zhang and Shasha(1989)]{zhang1989simple}
Kaizhong Zhang and Dennis Shasha.
\newblock Simple fast algorithms for the editing distance between trees and
  related problems.
\newblock \emph{SIAM journal on computing}, 18\penalty0 (6):\penalty0
  1245--1262, 1989.

\bibitem[Klein(1998)]{klein1998computing}
Philip~N Klein.
\newblock Computing the edit-distance between unrooted ordered trees.
\newblock In \emph{European Symposium on Algorithms}, pages 91--102. Springer,
  1998.

\bibitem[Demaine et~al.(2009)Demaine, Mozes, Rossman, and
  Weimann]{demaine2009optimal}
Erik~D Demaine, Shay Mozes, Benjamin Rossman, and Oren Weimann.
\newblock An optimal decomposition algorithm for tree edit distance.
\newblock \emph{ACM Transactions on Algorithms (TALG)}, 6\penalty0
  (1):\penalty0 1--19, 2009.

\bibitem[Hochsmann et~al.(2003)Hochsmann, Toller, Giegerich, and
  Kurtz]{hochsmann2003local}
Matthias Hochsmann, Thomas Toller, Robert Giegerich, and Stefan Kurtz.
\newblock Local similarity in rna secondary structures.
\newblock In \emph{Computational Systems Bioinformatics. CSB2003. Proceedings
  of the 2003 IEEE Bioinformatics Conference. CSB2003}, pages 159--168. IEEE,
  2003.

\bibitem[Quadrini et~al.(2019)Quadrini, Tesei, and
  Merelli]{quadrini2019algebraic}
Michela Quadrini, Luca Tesei, and Emanuela Merelli.
\newblock An algebraic language for rna pseudoknots comparison.
\newblock \emph{BMC bioinformatics}, 20\penalty0 (4):\penalty0 1--18, 2019.

\bibitem[Gellert et~al.(1962)Gellert, Lipsett, and Davies]{gellert1962helix}
Martin Gellert, Marie~N Lipsett, and David~R Davies.
\newblock Helix formation by guanylic acid.
\newblock \emph{Proceedings of the National Academy of Sciences}, 48\penalty0
  (12):\penalty0 2013--2018, 1962.

\bibitem[Matthews(1975)]{matthews1975comparison}
Brian~W Matthews.
\newblock Comparison of the predicted and observed secondary structure of t4
  phage lysozyme.
\newblock \emph{Biochimica et Biophysica Acta (BBA)-Protein Structure},
  405\penalty0 (2):\penalty0 442--451, 1975.

\bibitem[Chicco and Jurman(2020)]{chicco2020advantages}
Davide Chicco and Giuseppe Jurman.
\newblock The advantages of the matthews correlation coefficient (mcc) over f1
  score and accuracy in binary classification evaluation.
\newblock \emph{BMC genomics}, 21:\penalty0 1--13, 2020.

\bibitem[Wenter et~al.(2006)Wenter, F{\"u}rtig, Hainard, Schwalbe, and
  Pitsch]{wenter2006caged}
Philipp Wenter, Boris F{\"u}rtig, Alexandre Hainard, Harald Schwalbe, and
  Stefan Pitsch.
\newblock A caged uridine for the selective preparation of an rna fold and
  determination of its refolding kinetics by real-time nmr.
\newblock \emph{ChemBioChem}, 7\penalty0 (3):\penalty0 417--420, 2006.

\bibitem[Mathews(2019)]{mathews2019benchmark}
David~H Mathews.
\newblock How to benchmark rna secondary structure prediction accuracy.
\newblock \emph{Methods}, 162:\penalty0 60--67, 2019.

\bibitem[Woodson and Crothers(1987)]{woodson1987flucuatingpairs}
Sarah~A Woodson and Donald~M Crothers.
\newblock Proton nuclear magnetic resonance studies on bulge-containing dna
  oligonucleotides from a mutational hot-spot sequence.
\newblock \emph{Biochemistry}, 26\penalty0 (3):\penalty0 904--912, 1987.

\bibitem[Shervashidze et~al.(2011)Shervashidze, Schweitzer, Van~Leeuwen,
  Mehlhorn, and Borgwardt]{shervashidze2011weisfeiler}
Nino Shervashidze, Pascal Schweitzer, Erik~Jan Van~Leeuwen, Kurt Mehlhorn, and
  Karsten~M Borgwardt.
\newblock Weisfeiler-lehman graph kernels.
\newblock \emph{Journal of Machine Learning Research}, 12\penalty0 (9), 2011.

\bibitem[Karp and Rabin(1987)]{karp1987efficient}
Richard~M Karp and Michael~O Rabin.
\newblock Efficient randomized pattern-matching algorithms.
\newblock \emph{IBM journal of research and development}, 31\penalty0
  (2):\penalty0 249--260, 1987.

\bibitem[Hofacker et~al.(1994)Hofacker, Fontana, Stadler, Bonhoeffer, Tacker,
  and Schuster]{hofacker_1994}
Ivo Hofacker, Walter Fontana, Peter Stadler, Sebastian Bonhoeffer, Manfred
  Tacker, and Peter Schuster.
\newblock {F}ast {F}olding and {C}omparison of {RNA} {S}econdary {S}tructures.
\newblock \emph{Monatshefte fuer Chemie/Chemical Monthly}, 125:\penalty0
  167--188, 02 1994.

\bibitem[Do et~al.(2006)Do, Woods, and Batzoglou]{do2006contrafold}
Chuong~B Do, Daniel~A Woods, and Serafim Batzoglou.
\newblock Contrafold: Rna secondary structure prediction without physics-based
  models.
\newblock \emph{Bioinformatics}, 22\penalty0 (14):\penalty0 e90--e98, 2006.

\bibitem[Reuter and Mathews(2010)]{reuter2010rnastructure}
Jessica~S Reuter and David~H Mathews.
\newblock Rnastructure: software for rna secondary structure prediction and
  analysis.
\newblock \emph{BMC bioinformatics}, 11\penalty0 (1):\penalty0 1--9, 2010.

\bibitem[Sato et~al.(2011)Sato, Kato, Hamada, Akutsu, and Asai]{sato2011ipknot}
Kengo Sato, Yuki Kato, Michiaki Hamada, Tatsuya Akutsu, and Kiyoshi Asai.
\newblock Ipknot: fast and accurate prediction of rna secondary structures with
  pseudoknots using integer programming.
\newblock \emph{Bioinformatics}, 27\penalty0 (13):\penalty0 i85--i93, 2011.

\bibitem[Janssen and Giegerich(2015)]{janssen2015pkiss}
Stefan Janssen and Robert Giegerich.
\newblock The rna shapes studio.
\newblock \emph{Bioinformatics}, 31\penalty0 (3):\penalty0 423--425, 2015.

\bibitem[Huang et~al.(2019)Huang, Zhang, Deng, Zhao, Liu, Hendrix, and
  Mathews]{huang2019linearfold}
Liang Huang, He~Zhang, Dezhong Deng, Kai Zhao, Kaibo Liu, David~A Hendrix, and
  David~H Mathews.
\newblock Linearfold: linear-time approximate rna folding by 5'-to-3'dynamic
  programming and beam search.
\newblock \emph{Bioinformatics}, 35\penalty0 (14):\penalty0 i295--i304, 2019.

\bibitem[wwp(2019)]{wwpdb_2019}
Protein data bank: the single global archive for 3d macromolecular structure
  data.
\newblock \emph{Nucleic acids research}, 47\penalty0 (D1):\penalty0 D520--D528,
  2019.

\bibitem[Consortium(2020)]{rnacentral2021rnacentral}
RNAcentral Consortium.
\newblock {RNAcentral 2021: secondary structure integration, improved sequence
  search and new member databases}.
\newblock \emph{Nucleic Acids Research}, 49\penalty0 (D1):\penalty0 D212--D220,
  10 2020.
\newblock ISSN 0305-1048.
\newblock \doi{10.1093/nar/gkaa921}.

\bibitem[Roy et~al.(1990)Roy, Delling, Chen, Rosen, and
  Sonenberg]{roy1990bulge}
S~Roy, U~Delling, C-HRCA Chen, CA~Rosen, and N~Sonenberg.
\newblock A bulge structure in hiv-1 tar rna is required for tat binding and
  tat-mediated trans-activation.
\newblock \emph{Genes \& development}, 4\penalty0 (8):\penalty0 1365--1373,
  1990.

\bibitem[Kulinski et~al.(2003)Kulinski, Olejniczak, Huthoff, Bielecki,
  Pachulska-Wieczorek, Das, Berkhout, and Adamiak]{kulinski2003apical}
Tadeusz Kulinski, Mikolaj Olejniczak, Hendrik Huthoff, Lukasz Bielecki,
  Katarzyna Pachulska-Wieczorek, Atze~T Das, Ben Berkhout, and Ryszard~W
  Adamiak.
\newblock The apical loop of the hiv-1 tar rna hairpin is stabilized by a
  cross-loop base pair.
\newblock \emph{Journal of Biological Chemistry}, 278\penalty0 (40):\penalty0
  38892--38901, 2003.

\bibitem[Chen et~al.(2018)Chen, Feng, Yue, Bazdar, Mbonye, Zender, Harding,
  Bruggeman, Karn, Sieg, et~al.]{chen2018exosomes}
Lechuang Chen, Zhimin Feng, Hong Yue, Douglas Bazdar, Uri Mbonye, Chad Zender,
  Clifford~V Harding, Leslie Bruggeman, Jonathan Karn, Scott~F Sieg, et~al.
\newblock Exosomes derived from hiv-1-infected cells promote growth and
  progression of cancer via hiv tar rna.
\newblock \emph{Nature communications}, 9\penalty0 (1):\penalty0 4585, 2018.

\bibitem[Damke et~al.(2020)Damke, Melnikov, and
  H{\"u}llermeier]{damke2020novel}
Clemens Damke, Vitalik Melnikov, and Eyke H{\"u}llermeier.
\newblock A novel higher-order weisfeiler-lehman graph convolution.
\newblock In \emph{Asian Conference on Machine Learning}, pages 49--64. PMLR,
  2020.

\bibitem[Runge et~al.(2023)Runge, Franke, and Hutter]{runge2023towards}
Frederic Runge, J{\"o}rg~KH Franke, and Frank Hutter.
\newblock Towards automated design of riboswitches.
\newblock \emph{arXiv preprint arXiv:2307.08801}, 2023.

\bibitem[Runge et~al.(2019)Runge, Stoll, Falkner, and Hutter]{runge_2019}
Frederic Runge, Danny Stoll, Stefan Falkner, and Frank Hutter.
\newblock Learning to design {RNA}.
\newblock In \emph{International Conference on Learning Representations}, 2019.

\bibitem[Koodli et~al.(2021)Koodli, Rudolfs, Wayment-Steele, Designers, and
  Das]{koodli2021redesigning}
Rohan~V Koodli, Boris Rudolfs, Hannah~K Wayment-Steele, Eterna~Structure
  Designers, and Rhiju Das.
\newblock Redesigning the eterna100 for the vienna 2 folding engine.
\newblock \emph{bioRxiv}, pages 2021--08, 2021.

\bibitem[Anderson-Lee et~al.(2016)Anderson-Lee, Fisker, Kosaraju, Wu, Kong,
  Lee, Lee, Zada, Treuille, and Das]{anderson-lee_2016}
Jeff Anderson-Lee, Eli Fisker, Vineet Kosaraju, Michelle Wu, Justin Kong,
  Jeehyung Lee, Minjae Lee, Mathew Zada, Adrien Treuille, and Rhiju Das.
\newblock {P}rinciples for predicting {RNA} secondary structure design
  difficulty.
\newblock \emph{Journal of molecular biology}, 428\penalty0 (5):\penalty0
  748--757, 2016.

\bibitem[Ali et~al.(2023)Ali, Goyal, Jhunjhunwala, Mitra, Trant, and
  Sharma]{ali2023structural}
Zakir Ali, Ambika Goyal, Ayush Jhunjhunwala, Abhijit Mitra, John~F Trant, and
  Purshotam Sharma.
\newblock Structural and energetic features of base--base stacking contacts in
  rna.
\newblock \emph{Journal of Chemical Information and Modeling}, 63\penalty0
  (2):\penalty0 655--669, 2023.

\bibitem[Dethoff et~al.(2012)Dethoff, Chugh, Mustoe, and
  Al-Hashimi]{dethoff2012dynamicstructures}
Elizabeth~A Dethoff, Jeetender Chugh, Anthony~M Mustoe, and Hashim~M
  Al-Hashimi.
\newblock Functional complexity and regulation through rna dynamics.
\newblock \emph{Nature}, 482\penalty0 (7385):\penalty0 322--330, 2012.

\bibitem[Ganser et~al.(2019)Ganser, Kelly, Herschlag, and
  Al-Hashimi]{ganser2019rnadynamics}
Laura~R Ganser, Megan~L Kelly, Daniel Herschlag, and Hashim~M Al-Hashimi.
\newblock The roles of structural dynamics in the cellular functions of rnas.
\newblock \emph{Nature reviews Molecular cell biology}, 20\penalty0
  (8):\penalty0 474--489, 2019.

\bibitem[Wachsmuth et~al.(2012)Wachsmuth, Findeiß, Weissheimer, Stadler, and
  Mörl]{wachsmuth_2012}
Manja Wachsmuth, Sven Findeiß, Nadine Weissheimer, Peter~F. Stadler, and Mario
  Mörl.
\newblock {De novo design of a synthetic riboswitch that regulates
  transcription termination }.
\newblock \emph{Nucleic Acids Research}, 41\penalty0 (4):\penalty0 2541--2551,
  12 2012.
\newblock ISSN 0305-1048.

\end{thebibliography}

\newpage
\appendix

\section{Additional Experiments}
\label{app:add_exp}
In this section, we provide additional results for the comparison of the performance measures, F1 score, MCC, F1-shift, and the proposed Weisfeiler-Lehman graph kernel (WL). In Section~\ref{app:bulge_migration}, we show results for a simulated bulge migration experiment. Section~\ref{app:mutation} describes an experiment to analyze the influence of mutations on the scores of the performance measures.

\subsection{Analysis of Properties}
\label{app:bulge_migration}
\begin{figure*}[ht]
\vskip 0.2in
\begin{center}
        \centerline{%
        \includegraphics[width=\textwidth]{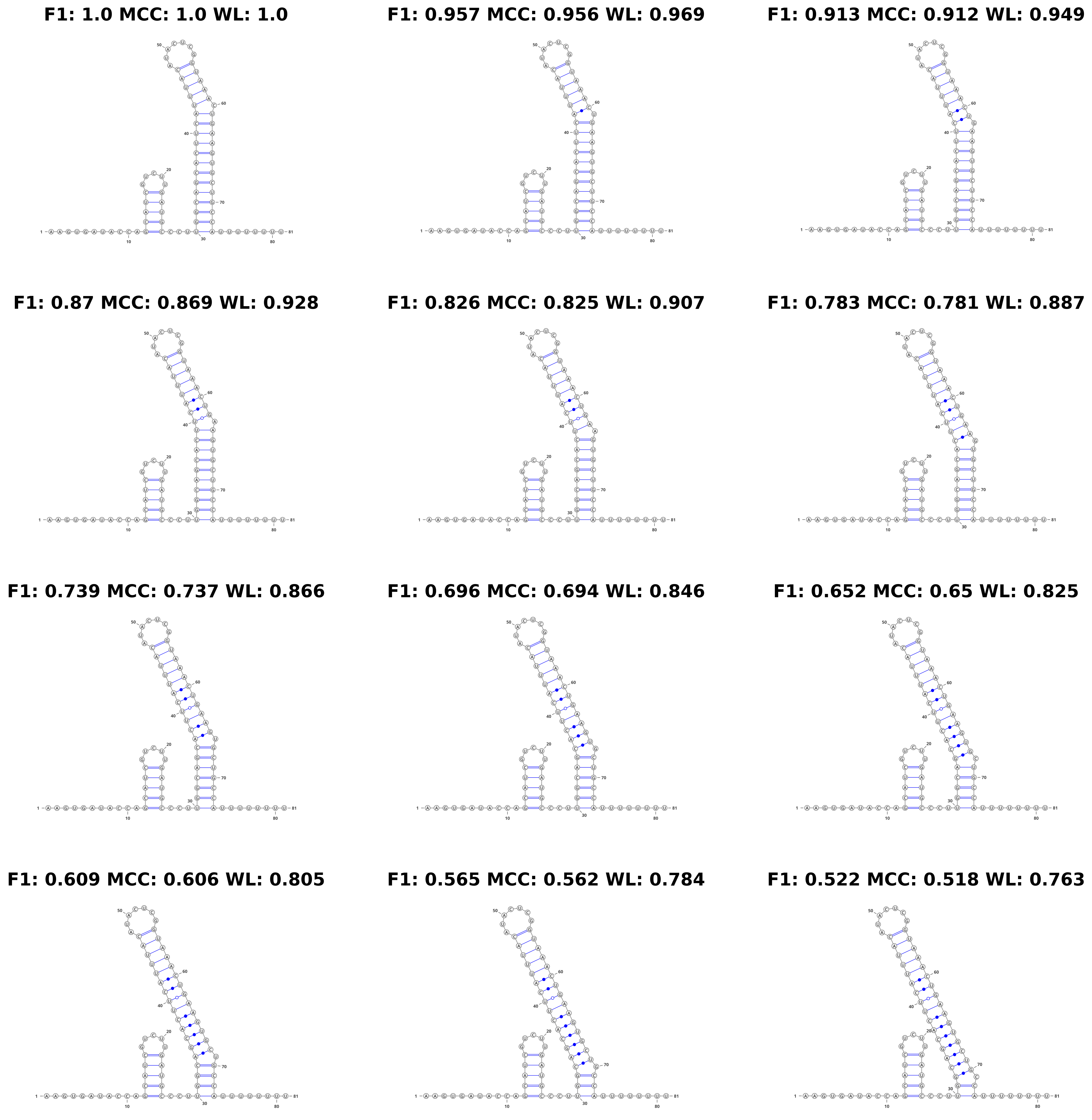}%
        }%}
\caption{
    \textbf{Bulge migration example.} We show an example of a simulated bulge migration process on a synthetic theophylline riboswitch construct RS3 proposed by \citet{wachsmuth_2012}. Top left shows the original construct. With each step, the bulge in the right stem is moving by one position.
	}	\label{fig:bulge_migration}
\end{center}
\vskip -0.2in
\end{figure*}

RNA structures are dynamic~\citep{dethoff2012dynamicstructures, ganser2019rnadynamics}. One example of such a dynamic behavior is the phenomenon of bulge migration~\citep{woodson1987flucuatingpairs}, where unpaired nucleotides on one strand of a stem move across the stem. For our analysis, we simulate such a migration event using a synthetic theophylline riboswitch construct, RS3, proposed by \citet{wachsmuth_2012}. The structure of RS3 contains a single nucleotide bulge at position 60 (see Figure~\ref{fig:bulge_migration}, top left) which we successively move by one position downstream in the stem. The results are shown in Figure~\ref{fig:bulge_migration}. Since F1-shift was introduced to capture exactly such shifts, the F1-shift score constantly stays at $1.0$ for all structures. However, the structures differ and not all bulges show a migration behaviour. Optimally, a performance measure, thus, should be capable of reflecting this in the scores. While the other measures all show reduced scores with increasing distance of the bulge from the original position, we observe that the WL shows the best balance since the decrease in the score is smoother and less drastically compared to the F1-score and the MCC.

\clearpage

\subsection{Application to Analysis}
\label{app:mutation}
\begin{figure*}[ht]
\vskip 0.2in
\begin{center}
        \centerline{%
        \includegraphics[width=.6\textwidth]{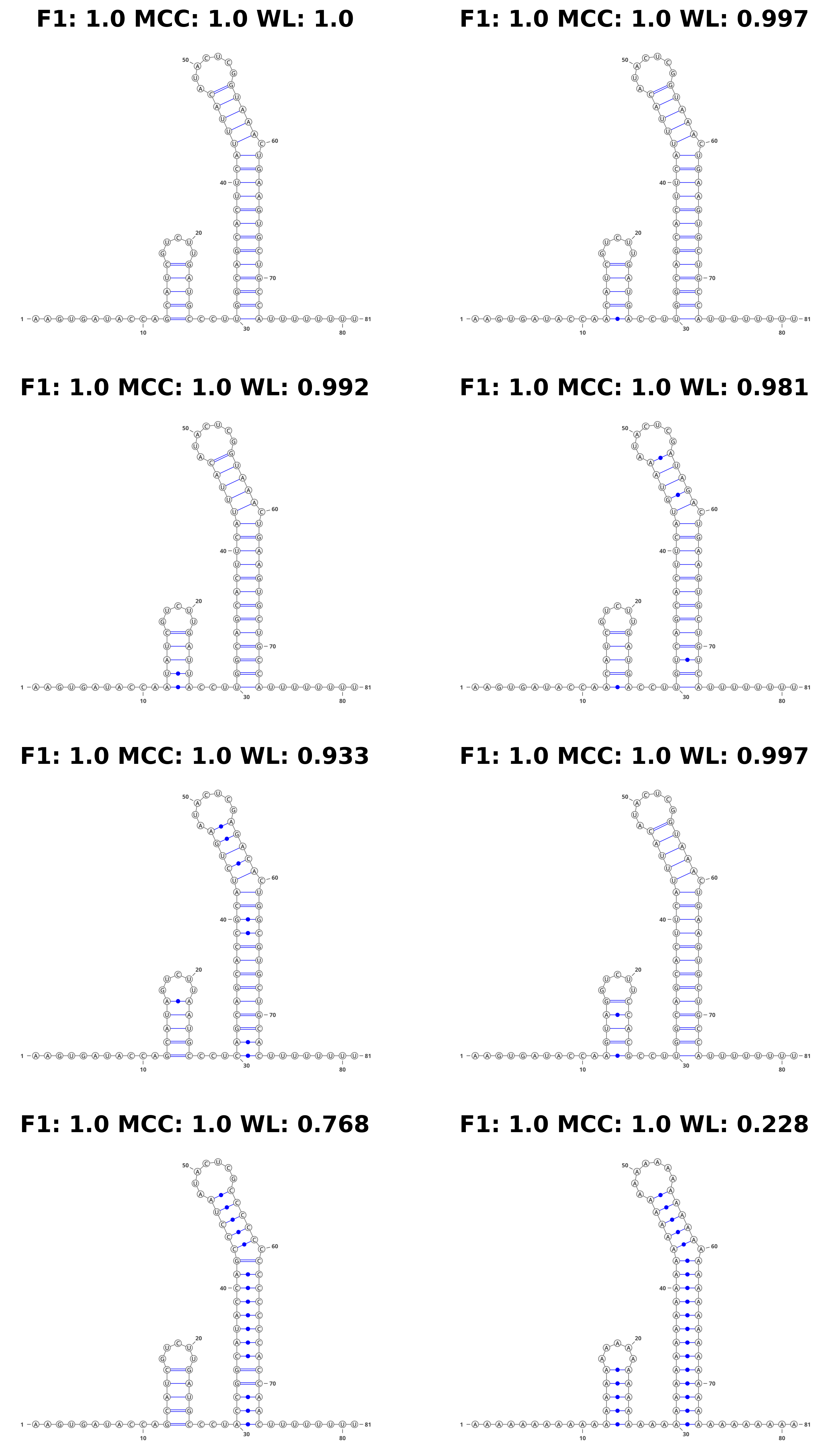}%
        }%}
\caption{
    \textbf{Mutation Example.} We show an example of a simulated mutation process on a synthetic theophylline riboswitch construct RS3 proposed by \citet{wachsmuth_2012}. Top left shows the original construct. With each step (left to right, top to bottom), we introduce the following mutations: 1 base pair (bp) mutated, 2 bp, 4 bp, 8 bp, entire first stem, entire second stem, entire sequence to 'A'.
	}	\label{fig:mutation}
\end{center}
\vskip -0.2in
\end{figure*}

In this section, we show the benefits of the property of WL to consider changes in the sequence additional to the structure for scoring the distance. 
We again use the theophylline riboswitch construct, RS3, and introduce mutations of base pairs.
Specifically, we either randomly change one, two, four, or eight base pairs, all base pairs of the first stem, all base pairs of the second stem, or replace the entire sequence with A's only. Figure~\ref{fig:mutation} shows the structures and the respective scores for the performance measures.
Except for WL, all performance measures are not capable of capturing changes in the sequence for their scores and the scores stay at $1.0$ for these measures since the structures remain unchanged. In contrast, the WL captures the mutation changes very well with decreasing scores depending on the number of mutations that we introduce. This property is a strong advantage of WL and could allow to use the WL kernel in evolutionary studies in the future. Further, besides capturing topological differences in RNA structures, the WL is generally capable of capturing changes in the base pair composition, which could allow for more fine-grained evaluation of structure prediction. 

\end{document}